\documentclass{article}

\usepackage[utf8]{inputenc}
\usepackage[T1]{fontenc}
\usepackage[english]{babel}
\usepackage{textcomp}
\usepackage{amsmath,amssymb}
\usepackage{lmodern}
\usepackage{graphicx}
\usepackage[dvipsnames,svgnames]{xcolor}
\usepackage{microtype}
\usepackage{hyperref} \hypersetup{pdfstartview=XYZ}

\usepackage{hyperref}

\usepackage{caption}
\usepackage{multirow}
\usepackage{supertabular}

\usepackage{tikz}
\usetikzlibrary{patterns}

\usepackage{authblk}

\title{Stark profiles modeling of radiation lines originating from atomic autoionizing states in dense plasmas, solid state matter under short intense XUV/X-ray free electron laser irradiation}

\author{Y.J. Aouad\footnote{PhD in Plasma Physics from UPMC "Université Pierre et Marie Curie" in the laboratory LULI "Laboratoire pour l'Utilisation des Lasers Intenses" - Ecole Polytechnique. Email: aouad852000@yahoo.fr}}
\affil{\footnotesize Theoretical Physics, Atomic Physics in Plasmas, France}

\begin{document}

\maketitle

\begin{abstract}

In the present paper we propose to update the standard calculation of Stark line profiles for autoionizing atomic states in dense plasmas, i.e. warm dense matter and strongly coupled plasmas. This is motivated by the importance of taking into account the effect of the electric microfield generated by the plasma charge constituents in the calculation of the atomic populations of autoionizing states which leads to the modification of the calculation of Stark line profiles in the frame of the standard theory. This is due to the properties of the autoionizing atomic states that are far from Boltzmann distribution even in dense plasmas bringing serious doubts on the statistical approach used in the standard theory of Stark profile calculations. We discuss the importance of the present analysis in view of the radiation emission originating from dense plasmas created by the interaction of the X-ray Free Electron Lasers (XFEL’s) with solid density matter.  
  
\end{abstract}

\tableofcontents
\bigskip

\section{Introduction}

\subsection*{XFEL’s interaction with a solid state matter: creation of warm dense matter and strongly coupled plasmas}
\hspace{2mm} In the past few years, X-ray free electron lasers (XFEL's) (installations: LCLS 2011, XFEL 2011, SACLA XFEL 2011)  offer for the first time to the scientific community the opportunity to study new regimes of matter \cite{Rosmej4, Rosmej1, Youcef5}. The XFEL's has outstanding properties: $10^{12}$ photons per pulse, photon energy in the XUV or X-ray range, short pulse duration of order of $10-100$ fs, high repetition frequency ($10-120$ Hz). These parameters have initiated an interest in different areas of research: planetary science, astrophysics, inertial-confinement fusion, high-energy density physics, and the study of exotic states of matter never created in laboratories so far. 
\\
\\
Because the critical density of the XFEL’s is larger than the solid density, the absorption of the radiation energy of the laser is homogeneous and proceeds into the volume of the solid. This occurs on a short time scale when the matter is in its solid-state density allowing isochoric heating of the solid. The energy and intensity (of order of $10^{+16}$ W/cm$^{2}$) of the XFEL's may induce the intern shell photoionization of most atoms of the crystal and because the pulse duration of the XFEL's is of the order of the autoionization rate, hollow ions are formed and the solid can be transformed into an exotic state of matter: a hollow crystal \cite{Rosmej1, Galtier1}. The mechanisms of evolution of the hollow crystal to warm dense matter to strongly coupled plasma are not well understood.     

\subsection*{X-ray emission originating from autoionizing atomic states and hollow ion configurations: time resolved spectroscopy for the investigation of warm dense matter and strongly coupled plasmas}
\hspace{2mm}  Nevertheless, high-resolution XUV or X-ray spectroscopy (gratings and Bragg crystals) is an important method that enables the study of the disintegration of crystalline order and material heating as has been demonstrated \cite{Galtier1} with the FLASH free electron laser in Hamburg where electron densities $n_e$ of about $10^{+22} - 10^{+23}$ cm$^{-3}$ and electron temperatures $T_e \approx 30$ eV have been deduced from the analysis of intensities of $M-L$ and $L-L$ transitions in multiple excited aluminum ions. This analysis also allowed researchers to describe the evolution of the heated sample (disintegration of crystalline order).
\\
\\
This is based on the properties of the XFEL’s itself where the creation of multi-excited states and hollow ion configurations by direct photoionization of the $K$ and $L$ atomic shells is possible as the energy of the XFEL's photons allows internal shells ionizations \cite{Rosmej5, Rosmej2}. Autoionizing configurations provide an outstanding tool for the study of dense plasma regimes \cite{Youcef1, Youcef5}. This concerns both, intensities of the emitted lines that are related to populations of autoionizing states and line broadening. In \cite{Galtier2}, the complex line broadening of $L-L$ intrashell transitions were proposed to make use of the information contained in the contour shape of the $M-L$ and $L-L$ transitions of Al III and Al IV corresponding to the emission of the type $K^2L^7M^m \rightarrow K^2L^8M^{m-1} + h\nu$ and $K^22s^12p^6M^m \rightarrow K^22s^22p^5M^m + h\nu$ with $m = 2$ and $1$. In \cite{Galtier2} detailed calculations were carried out with the PPP code \cite{Calisti1, Calisti2}, which provides fast and accurate line profiles in plasmas for very complex transition arrays. PPP accounts for all the main mechanisms of line broadening: lifetime broadening due to spontaneous emission, and Stark broadening. 
\\
\\
However, the calculation of Stark broadening in the PPP code is based under the assumption that the populations are in thermodynamical equilibrium in dense plasmas. The situation is more complicated for the atomic autoionizing configurations that are characterized by a high autoionization rates of the order of $\approx 10^{+13}-10^{+15}$ s$^{-1}$ \cite{Rosmej1, Youcef1, Youcef5}. Therefore, local thermodynamical equilibrium populations can only be achieved for electron densities above solid density and non local thermodynamical equilibrium populations have to be considered for almost all parameters of practical interest. Also, in dense plasma regimes the effect of the plasma electric microfield has to be taken into account in the calculation of the atomic populations by the use of the density matrix kinetics model instead of the standard collisional-radiative one \cite{Youcef1, Youcef5}.   
\\
\\
In the present paper, we investigate the theoretical background of the use of the atomic density matrix kinetics in the calculation of line broadening of autoionizing atomic states in dense plasmas. We give the analytical formula that introduce the field dependent atomic populations in the calculation of the spectral line shapes in the case of the static ion microfield model and the frequency fluctuation model (FFM) \cite{Calisti2}.  


\section{Spectral line shapes: standard theory and the code PPP}
\hspace{2mm} Line shape analysis is an important tool for the investigation of dense plasma regimes, as it gives information on the physical processes involved in the line formation. The modeling of line broadening from neutral or charged emitters has been in perpetual development and remains a keystone in plasma spectroscopy \cite{Griem1}. In the calculation of a line shape, Stark broadening is the most computationally challenging contribution, since the main difficulty is to properly characterize the emitter environment. It involves a complex combination of atomic physics, statistical mechanics and detailed plasma physics \cite{Griem2}.
\\
\\
In the dipole approximation, the line shape $I(\omega)$ is related to the Fourier-transforms of the autocorrelation function $C(t)$ of the radiator dipole operator $\hat{\vec{d}}$ and is given by: 
\\
\begin{equation}\label{Line_omega}
I(\omega)= \frac{1}{\pi}Re\int_{0}^{+\infty}dt \ e^{i\omega t} \ C(t)
\end{equation}
\\ 
where the autocorrelation function $C(t)$ of the radiator dipole operator $\hat{\vec{d}}$ is given by:
\\
\begin{equation}\label{autocorrelation_dipole}
C(t)= <<\hat{\vec{d}} \ ^\dag \mid \hat{U}(t) \mid \hat{\vec{d}} \ \rho_{eq} >>
\end{equation}
\\ 
where the double bra and ket are defined in the Liouville space  (the Liouville space is isomorphic to the direct product of the Hilbert space with its dual space corresponding to the quantum physical system in question). In Eq.\ref{autocorrelation_dipole} , $\rho_{eq}$ is the thermodynamical equilibrium density matrix of the emitter and $\hat{U}(t)=\{\hat{U}_l(t)\}_{l\in \{F\}}$ is the bath averaged evolution operator of the emitter where $l$ belongs to a measurable functional space $\{F\}$ which provides a statistical method for the calculation of average quantities. $\hat{U}_l(t)$ is solution of the following stochastic Liouville equation (SLE):
 \\
\begin{equation}\label{U_SLE}
\frac{d\hat{U}_l(t)}{dt}=-iL_l \cdot \hat{U}_l(t)
\end{equation}
\\ 
with the condition $\hat{U}_l(0)= \mathbb{I}_d $, the identity operator. $L_l $ is the Liouvillian operator of the emitter in the plasma bath. This operator is composed of the free emitter part $L_0$ of the unperturbed atom and a random perturbation part describing the interaction of the emitter with the plasma bath that includes the electronic contribution $L_{e,l}$ as well as the contribution of ions $L_{i,l}$. 
\\
\begin{equation}\label{Liouvillian_l}
L_l = L_0+L_{i,l}+L_{e,l}
\end{equation}
\\
In the PPP code, the random perturbation of the bath is given in the frame of the standard theory. Ions and electrons are treated in different ways due to their mass difference. The fast electrons are assumed to perturb the emitter by means of collisions, this part is introduced in the impact approximation by the non-Hermitian electronic operator $\Phi$. The impact approximation is based on the assumption of short and rare binary collisions occurring between the emitter and the electronic perturbers and the mean time between two binary collisions is much longer than the duration of the collision. And the slow ions contribution is treated in the static limit in which the ions electric microfield acting on the emitter is taken to be constant during the radiative process. This part is taken in the dipole interaction approximation where the ions electric microfield $\vec{E}$ is introduced and characterized by a stationary-field probability density distribution function $Q(\vec{E})$. Then, the Liouvillian operator $L(\vec{E}) $ is given by: 
\\
\begin{equation}\label{Liouvillian}
L(\vec{E}) = L_0-\frac{1}{\hbar}\hat{\vec{D}}\cdot\vec{E}-i\Phi
\end{equation}

\begin{equation}\label{Liouvillian_0}
L_0 = \hat{H}_0\otimes\mathbb{I}_d-\mathbb{I}_d\otimes \hat{H}_0^{*}
\end{equation}

\begin{equation}\label{dipole_liouville}
\hat{\vec{D}}=\hat{\vec{d}}\otimes\mathbb{I}_d-\mathbb{I}_d\otimes \hat{\vec{d}}^{*}
\end{equation}
\\
where in Eq.\ref{Liouvillian_0}, $\hat{H}_0$ is the unperturbed Hamiltonian of the emitter. In Eq.\ref{dipole_liouville}, $\hat{\vec{D}}$ is the dipole operator of the emitter written in Liouville space and $\hat{\vec{d}}$ is the dipole operator of the emitter written in Hilbert space. These assumptions lead to a spectral line shape function that can be written as a sum of rational fractions or generalized Lorentzian spectral components of the line characterized by a complex frequency and a complex intensity due to the fact that the electronic collision operator is non-Hermitian. It is to note that the PPP code can also take into account the dynamics of ions in the frame of the frequency fluctuation model (FFM) (see below).

\subsection{General discussion of the standard theory: introduction of the field dependent density matrix}
\hspace{2mm}  As seen in Eq.\ref{autocorrelation_dipole}, the standard theory of the spectral line shapes calculation and the code PPP are based on the assumption of a thermodynamical equilibrium density matrix $\rho_{eq}$:
\\
\begin{equation}\label{Eq_DM}
\rho_{eq}=\frac{e^{-\beta \hat{H}_{0}}}{\text{Tr}(e^{-\beta \hat{H}_{0}})}
\end{equation}
\\ 
where, $\text{Tr}$ is the trace calculation operation and $\beta=1/(k_B T_e)$, $k_B$ is the Boltzmann constant and $T_e$ is the electronic temperature. However, the atomic populations of autoionizing states are far from the Boltzmann distribution even in dense plasmas due to their huge autoionization rate \cite{Youcef1}. In this case, non-equilibrium populations have to replace the equilibrium ones in Eq.\ref{autocorrelation_dipole}. At the same time, in dense plasmas the electric microfield generated by the ions on the emitter has to be considered in the calculation of the atomic populations of autoionizing states \cite{Youcef5}. Thus for populations of autoionizing atomic states, the standard non-equilibrium collisional-radiative model must be replaced by the atomic density matrix kinetics model which leads to a plasma electric microfield dependent atomic populations. This procedure leads to the changing of the field average calculation in Eq.\ref{autocorrelation_dipole} in which the average over the electric microfield distribution function $Q(\vec{E})$ in the static limit enters $\hat{U}(t)$ only and excludes the density matrix of the emitter that changes from $\rho_{eq}$ to $\rho(\vec{E})$. In the next section, we describe the system of equations that leads to a field dependent density matrix $\rho(\vec{E})$ in dense plasmas.


\section{Field dependent atomic kinetics in dense plasmas: density matrix kinetic equation in plasmas}
\hspace{2mm} The effect of the plasma on the atomic kinetics concerns both, the non dissipative Hamiltonian  mixing of the atomic levels by the quasi-static ion electric microfield and the dissipative time evolution part induced by the multiple atomic processes in the plasma \cite{Youcef1, Youcef5}. The latter part is often introduced in the frame of the collisional radiative kinetics model where in the density matrix case the difference is given by the relaxation of the non-diagonal density matrix elements also called coherences. The matrix elements representation $\rho_{\alpha\beta}$ of the density operator $\rho$ for the two unperturbed atomic states $\mid \alpha >$ and $\mid \beta >$ writes:
\\
\begin{equation}\label{rho_jk}
\rho_{\alpha\beta}= < \alpha \mid \rho \mid \beta >
\end{equation}
\\ 
The time evolution equation of $\rho_{\alpha\beta}$ is given by:
\\
\begin{multline}\label{ATM_tev_dm}
\frac{\partial \rho_{\alpha\beta}}{\partial t} = - \frac{i}{\hbar} \left( E_\alpha - E_\beta \right) \rho_{\alpha\beta} -\frac{i}{\hbar} \sum_{\gamma} \left(\hat{V}(\vec{E})_{\alpha\gamma} \times \rho_{\gamma\beta} - \rho_{\alpha\gamma} \times \hat{V}(\vec{E})_{\gamma\beta} \right) -\\
\frac{1}{2} \left( \sum_{\gamma} \left(W_{\alpha\gamma} + W_{\beta\gamma} \right) \right) \rho_{\alpha\beta} + \delta_{\alpha\beta} \sum_{\gamma} W_{\gamma\beta} \times \rho_{\gamma\gamma}
\end{multline}
\\ 
where $\hbar$ is the reduced Planck constant and the first term of the right hand side of Eq.\ref{ATM_tev_dm} includes  $\hat{H}_{0}$ the unperturbed Hamiltonian of the emitting ion, $\hat{V}$ the interaction of the emitting ion with the static electric microfield $\vec{E}$ of the surrounding ions in the dipole approximation. More precisely, the use of the static approximation is justified by the fact that the relaxation time of the density matrix elements is smaller than the inverse plasma frequency associated to the ions. Denoting the dipole atomic moment of the emitting ion $\hat{\vec{d}}$, the interaction potential energy $\hat{V}$ is given by: 
\\
\begin{equation}\label{V_op}
\hat{V}(\vec{E})=-\hat{\vec{d}} \cdot \vec{E}
\end{equation}
\\
The dissipative part in Eq.\ref{ATM_tev_dm} that is expressed in terms of the standard transition rates $W_{\alpha\beta}$ from the atomic state $\mid \alpha >$ to the atomic state $\mid \beta >$ includes all the collisionnal and radiative relaxation processes that populate and depopulate the elements of the density matrix, i.e., spontaneous emission rate $A$ and autoionization rate $\Gamma$, collisionnal relaxation between levels that are mixed by the electric microfield $\vec{E}$ of ions and so on. It is to note that the solution density matrix of Eq.\ref{ATM_tev_dm} is field dependent: 
\\
\begin{equation}\label{rho_E}
\rho \equiv \rho(\vec{E})
\end{equation}
\\
If one is interested by the calculation of the atomic population of the atomic state $\mid \alpha >$, the final result must be averaged over the plasma electric microfield distribution function $Q(\vec{E})$:
\\
\begin{equation}\label{rho_average}
<\rho_{\alpha\alpha}>=\int d\vec{E} \ Q(\vec{E}) \rho_{\alpha\alpha}(\vec{E})
\end{equation}
\\
However, the calculation of the spectral line shape uses the field dependent density matrix $\rho(\vec{E})$ before the field average procedure. The average calculation includes other terms in the integrand as described in the next section.  


\section{Modification of the spectral line shape calculations in dense plasmas: Stark profiles of dielectronic satellites in dense plasmas}    
\hspace{2mm} The introduction of the atomic density matrix kinetics in the calculation of the spectroscopic properties of autoionizing atomic states concerns both, the atomic populations and the spectral line shapes \cite{Youcef1, Youcef5}. The modification of the spectral line shapes calculation by taking into account the plasma electric microfield dependent atomic density matrix $\rho(\vec{E})$ Eq.\ref{ATM_tev_dm}-\ref{rho_E} in the standard theory is discussed below. Firstly in the case of the static ion microfield limit and secondly in the frame of the frequency fluctuation model (FFM). Both of these models are implemented in the PPP code. 

\subsection{The case of static ion microfield limit}
 In this context, the modification of the spectral line shape function $I(\omega)$ Eq.\ref{Line_omega} enters in the calculation of the autocorrelation function $C(t)$ Eq.\ref{autocorrelation_dipole}. The modification consists firstly by introducing the plasma electric microfield dependent atomic density matrix $\rho(\vec{E})$ Eq.\ref{ATM_tev_dm}-\ref{rho_E} and secondly by changing the average procedure over the static plasma electric microfield distribution function $Q(\vec{E})$. The algebraic derivations lead to the following final expression for $I(\omega)$:
\\
\begin{equation}\label{SLS}
I(\omega) = - \frac{1}{\pi}\text{Im}\sum_{k=1}^{N}\int d\vec{E} \ Q(\vec{E}) \frac{\left( \hat{\vec{d}}^\dagger M(\vec{E}) \right)_{kk}\left( M^{-1}(\vec{E}) \rho(\vec{E}) \hat{\vec{d}} \right)_{kk}}{\left( \omega - f_{k}(\vec{E}) \right)-i\gamma_{k}(\vec{E})}
\end{equation}
\\
where, $N$ is the number of atomic levels entering in the calculation of the trace in the evaluation of the dipole autocorrelation function, $\hat{\vec{d}}$ is the dipole operator of the emitting ion, $M(\vec{E})$ is the similarity diagonalisation matrix of the field dependent Liouvillian operator $L(\vec{E})$ Eq.\ref{Liouvillian}, $(f_k({\vec{E}})+i\gamma_k({\vec{E}}))$ is the $k^{th}$ complex electric field dependent diagonal element of the Liouvillian Eq.\ref{Liouvillian} so that $f_k(\vec{E})$ is the Stark shifted frequency and $\gamma_k(\vec{E})$ is related to the impact broadening electronic operator $\Phi$ Eq.\ref{Liouvillian} and represents the width of the emitted line.


\subsection{Line shapes including the dynamics of ions in the frame of frequency fluctuation model (FFM)}
\hspace{2mm} In addition to the static ion electric microfield limit, the code PPP can take into account the dynamical effect of ions during the radiative emission and this is implemented in the frequency fluctuation model (FFM) \cite{Calisti2}. In this case also, the plasma electric microfield dependent atomic density matrix $\rho(\vec{E})$ Eq.\ref{ATM_tev_dm}-\ref{rho_E} must be introduced for the calculation of the line profile of autoionizing atomic states in dense plasma regimes. 
\\
\\
In the FFM model, the perturbed emitter by a time-dependent plasma ion microfield behaves like a set of field-dressed two-level transitions subject to a collision-type mixing process. More precisely, this model is based on the assumption that the microfield fluctuations induce an emitted frequency fluctuations. In the sens that the dynamic of the field amplitude is considered as a stochastic process involving a transition between its values at a jumping frequency $\nu \sim (v_{thi}/a)$ ($v_{thi}$ is the ion thermal velocity, $a=\left( \frac{4\pi}{3}n_i\right) ^{-1/3}$ is the ion spehere radius where $n_i$ is the ion density). This mecanism induces a stationnary stochastic Markov process among the Stark components of the emitted frequencies of the ion. The Stark components are derived from the approximation of a static ions so that each Stark component is represented by: one value of the electric microfield amplitude, field dependent frequency and a field dependent complex intensity. The expression of the intensity is given by attributing to each Stark component an index $i$ so that the sum rules are now performed over the Stark components. The line profile is given by:
\\
\begin{equation}\label{line_profile}
I(\omega) = \text{Re} \sum_{i,j} \left( \omega - f - i\gamma - iW \right)_{ij}^{-1} \left( a_i + i c_i \right)  
\end{equation} 

\begin{equation}\label{intensity_complex}
a_i + i c_i = -4 E^2 Q(\vec{E}) \left( \hat{\vec{d}}^\dagger M(\vec{E}) \right)_{ii}\left( M^{-1}(\vec{E}) \rho(\vec{E}) \hat{\vec{d}} \right)_{ii}
\end{equation}  
\\
where in Eq.\ref{line_profile}, W is proportional to the field stochastic jumping frequency $\nu$:
\\
\begin{equation}\label{W_field}
W_{ij} = -\nu \ p_i
\end{equation}

\begin{equation}\label{W_field_bis}
W_{ii} = \nu(1-p_i)
\end{equation}

\begin{equation}\label{p_i}
p_i = \frac{a_i}{\sum_j a_j}
\end{equation}
\\
where, $p_i$ represents the probability of the $i^{th}$ Stark component. As in the case of the static ion microfield limit, the plasma electric microfield dependent atomic density matrix $\rho(\vec{E})$ Eq.\ref{ATM_tev_dm}-\ref{rho_E} enters in Eq.\ref{intensity_complex} for the calculation of the line profile function $I(\omega)$ Eq.\ref{line_profile}.
\\
\\
In the next section, two expressions of the plasma electric microfield distribution function $Q(\vec{E})$ are given, namely the Holtsmark distribution function \cite{Holstmark1} and a fitted analytical expression that takes into account the ion correlations and the electron screening effect of the plasma. 


\section{Field distribution functions}
\hspace{2mm} In order to perform the average calculation over the microfield values, the microfield distribution function $Q(\vec{E})$ is needed. Taking into account the isotropy condition, the isotropic plasma electric microfield distribution function $W(E)$ is considered and reads:
\\
\begin{equation}\label{distribution_function}
W(E) = 4 \pi E^2 Q(\vec{E})
\end{equation}
\\
where $E$ is the electric microfield amplitude. For the representation of the distribution function, it is more convenient to introduce the dimensionless field $\beta$ Eq.\ref{beta} and the distribution function $P(\beta)$ Eq.\ref{P_beta}:
\\
\begin{equation}\label{beta}
\beta = \left( \frac{a^{2}}{Ze}\right) E
\end{equation} 
\\
\begin{equation}\label{P_beta}
P(\beta) \ d\beta = W(E) \ dE
\end{equation}
\\
where, $Ze$ is the ion charge, $a=\left( \frac{4\pi}{3}n_i\right) ^{-1/3}$ is the ion sphere radius, $n_i$ is the ion density. In the following we give two representations of the distribution function $P(\beta)$.


\subsection{Holtsmark electric microfield distribution function}
\hspace{2mm} This approximation is valid in the high temperature limit when we can neglect the particle correlations due to their Coulomb interaction by assuming that the ions are not correlated and that the electron screening is negligible so that the coulomb coupling parameter $\Gamma$ Eq.\ref{Coupling_parameter} is close to zero:  
\\
\begin{equation}\label{Coupling_parameter}
\Gamma = \frac{(Ze)^{2}}{ak_BT}
\end{equation}
\\
where, $T$ is the temperature, $k_B$ is the Boltzmann constant. The Holtsmark distribution function $P_H(\beta)$ \cite{Holstmark1} is given by:
\\
\begin{equation}\label{Holtsmark}
P_H(\beta) = \frac{2\beta}{\pi}\int_0^\infty x \ exp(-x^{3/2}) \ sin(\beta x) \ dx
\end{equation}
\\
In Figure.\ref{Distribution_fig}, the Holtsmark distribution function is represented.


\subsection{Fitted expression for the electric microfield distribution function}
\hspace{2mm} To take into account the ion correlations and the electron screening in the microfield distribution function, we can find in the literature some analytical expressions. As an example, we present the following expression \cite{Potekhin1}:
\\
\begin{equation}\label{fiited_formula}
P(\beta) \approx \frac{\beta^2}{S_N}\left[ Ae^{-a\beta^\alpha} + Be^{-b \beta^{\gamma}} + \frac{e^{-\Gamma \beta^{1/2}}}{1+c\beta^{9/2}} \right] 
\end{equation}
\\
where $S_N$ is the normalisation constant:
\\ 
\begin{equation}\label{normalisation_cst}
S_N = A\frac{\Gamma(3/\alpha)}{\alpha a^{3/2}} + B\frac{\Gamma(3/\gamma)}{\gamma b^{3/\gamma}} + \Gamma^{-6}F(c/\Gamma^{6})
\end{equation}
\\
where $\Gamma(3/\alpha)$ and $\Gamma(3/\gamma)$ are the Gamma-function values which are easily calculated (e.g., \cite{Press1}), and $F(y)$ is given by:
\\
\begin{equation}\label{F_y}
F(y) = \int_0^{\infty} \frac{x^{2}e^{-\sqrt{x}}}{1+yx^{9/2}} dx
\end{equation} 
\\
where in Eq.\ref{fiited_formula} and Eq.\ref{normalisation_cst} the coefficients $A$, $a$, $\alpha$, $B$, $b$ and $\gamma$ are functions of $\Gamma$ Eq.\ref{Coupling_parameter} and the electron screening parameter $s \sim (a/\lambda_{De})$. In Figure.\ref{Distribution_fig}, the distribution function Eq.\ref{fiited_formula} corresponding to $s=0$ and $\Gamma = 10^{-7}$ is represented and compared to the Holtsmark distribution function. These values lead to a fitted distribution close to the Holtsmark function as seen in Figure.\ref{Distribution_fig}. 


\section{Conclusion}
\hspace{2mm} In this paper we have motivated the introduction of the plasma electric microfield dependent atomic density matrix for the calculation of populations and line shapes of atomic autoionzing states in dense plasma regimes, i.e, warm dense matter (WDM) and strongly coupled plasmas (DSCP). We presented the formalism of the line shapes calculation in the frame of the standard theory and the code PPP. The plasma ion static electric microfield limit is presented as well as the frequency fluctuation model (FFM). The general formulation of the atomic density matrix applied to atomic kinetics in dense plasmas is presented. Analytical expressions of line profile functions using the plasma electric microfield dependent atomic density matrix are discussed and presented. This method could allow the study of the influence of the correlations between ions on the atomic populations in the warm dense matter (WDM) and the strongly coupled plasmas (DSCP). 


\begin{center}
\includegraphics[scale=0.5]{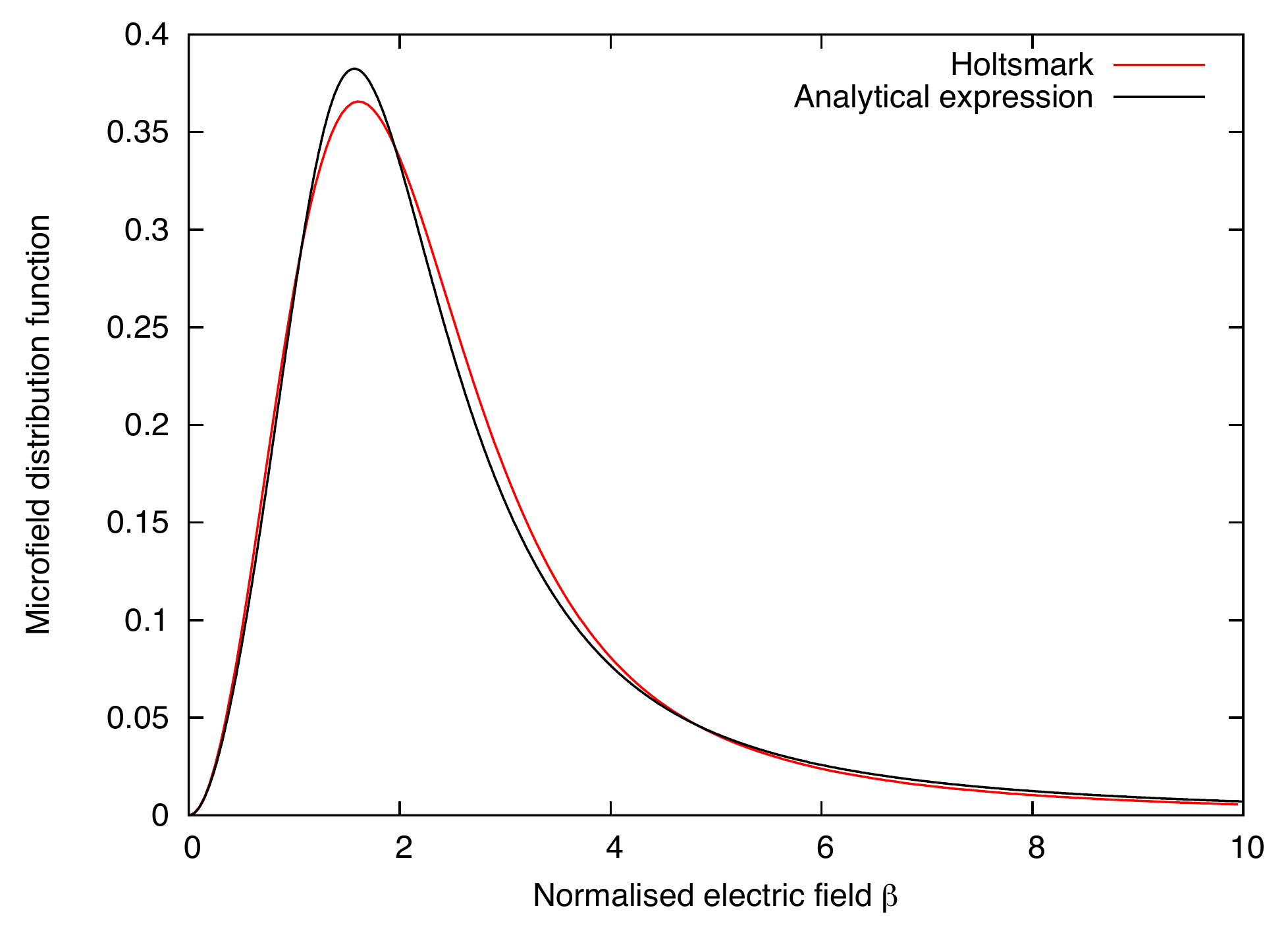}
\captionof{figure}{Field distribution function calculated using the Holtsmark approximation (red curve) and the analytical expression Eq.\ref{fiited_formula} (black curve).}\label{Distribution_fig}
\end{center}
     


\end{document}